\documentclass[useAMS,usenatbib]{mn2e}
\usepackage{amsmath,amssymb}
\usepackage{url}
\usepackage{graphicx}

\def\eg{{\it e.g.,}}
\def\HI{H{\small I}}
\def\d{{\mathrm d}}

\title[The Smith Cloud and its dark matter halo]{The Smith Cloud and its dark matter halo: Survival of a Galactic disc passage}
\author[Nichols et al.]{Matthew Nichols$^{1}$\thanks{E-mail:matthew.nichols@epfl.ch},
Nestor Mirabal$^{2,3}$\thanks{E-mail:mirabal@gae.ucm.es}, Oscar Agertz$^{4}$, 
Felix J. Lockman$^{5}$,\and Joss Bland-Hawthorn$^{6}$\\
$^{1}$ Laboratoire d'Astrophysique, \'Ecole Polytechnique F\'ed\'erale de Lausanne (EPFL), Observatoire de Sauverny, 1290 Versoix, Switzerland\\
$^{2}$ Ram\'on y Cajal Fellow\\
$^{3}$ Dpto. de F\'isica At\'omica,
Molecular y Nuclear, Universidad Complutense de
Madrid, Spain\\
$^{4}$ Department of Physics, University of Surrey, Guildford GU2 7XH,
United Kingdom\\
$^{5}$ National Radio Astronomy Observatory,  P.O. Box 2, Green Bank, 
\textsc{}WV 24944, United States\\
$^{6}$ Sydney Institute for Astronomy, School of Physics,
The University of Sydney, NSW 2006, Australia\\
}

\begin{document}

\date{}

\pagerange{\pageref{firstpage}--\pageref{lastpage}} \pubyear{2013}

\maketitle

\label{firstpage}

\begin{abstract}
The current velocity of the Smith Cloud indicates that it has undergone at least one passage of the Galactic disc.
Using hydrodynamic simulations we examine the present day structure of the Smith Cloud. 
We find that a dark matter supported cloud is able to reproduce the observed present day neutral hydrogen mass, column density distribution and morphology. In this case the dark matter halo becomes elongated, owing to the tidal interaction with the Galactic disc. Clouds in models neglecting dark matter confinement are destroyed upon disc passage, unless the initial cloud mass is well in excess of what is observed today.
We then determine integrated flux upper limits to the gamma-ray emission around  such a hypothesised dark matter core in the Smith Cloud.  No statistically significant core or
extended gamma-ray emission are detected down to a 95\% confidence level 
upper limit of 1.4 $\times 10^{-10}$ ph cm$^{-2}$ s$^{-1}$ in the 1--300 GeV energy range. 
For the derived distance of 12.4 kpc, the {\it Fermi} upper limits set 
the first tentative constraints on the dark matter cross
sections annihilating into $\tau^{+}{\tau}^{-}$ and $b\bar{b}$ for a high-velocity cloud.
\end{abstract}

\begin{keywords}
(cosmology:) dark matter -- gamma-rays: general -- ISM: clouds --
ISM: individual objects: Smith cloud  -- Galaxy: halo
\end{keywords}

\section{Introduction}

The mapping of dark matter substructure around the Galaxy is crucial to understanding how our own Milky Way was assembled over cosmic time.
Numerical simulations of the concordant cosmology, dark-energy and cold dark-matter ($\Lambda$CDM), predict a multitude of dark-matter subhaloes going down in mass to approximately $10^{-4}~M_\odot$ \citep[\eg][]{Klypin1999,Moore1999,Diemand2007,Springel2008}.
On the detectable scale of dwarf galaxies, this discrepancy approaches an order of magnitude between the observed dwarf galaxies and the number of predicted subhaloes \citep{Mateo1998,Weinberg2013}.
Recent discoveries of ultra-faint dwarf galaxies go someway towards filling this gap \citep[\eg][]{Willman2005,Belokurov2007}.
Unfortunately, due to the minuscule stellar populations, finding ultra faint dwarfs continues to be challenging and the exact number of them are unknown.
The low stellar content of them however, makes them potentially excellent objects to search for dark matter annihilation signals \citep{Charbonnier2011}.
So far, no significant detections have been made, although upper limits have been placed on the properties of dark matter \citep[\eg][]{dwarfs,Natarajan2013,Fermi2013}.

In addition to dwarf galaxies, the Galactic halo contains a large number of intriguing \HI{} substructures in the form of high velocity clouds (HVCs) \citep{Wakker1997}.
The suggestion of dark matter surrounding a population of these HVCs has been around for over a decade \citep[\eg][]{Blitz1999,Quilis2001}.
Subsequent investigation has determined that any such population is likely to be small in comparison to dark matter free HVCs, both around our own Galaxy \citep{Saul2012} and also around other galaxies \citep{Chynoweth2011}.
Near the disc some of the HVCs are likely to arise through a galactic fountain, where numerous supernova launch gas from the disc of the Galaxy \citep{Bregman1980}, however, this process is likely to produce small, intermediate velocity clouds over HVCs \citep{Ford2010}.
Many HVCs that lie within 50 kpc of the Milky Way likely had their origin in the Magellanic Stream and thus arose from the LMC and SMC \citep{Putman2004}.
Extragalactic HVCs are seen at a wide range of projected distances, often exceeding $150$~kpc from the nearest galaxy, are likely to be clumps of pristine gas infalling for the first time and possess a phase-space distribution that is incompatible with the expected dark matter substructure \citep{Chynoweth2011}.
Even at low projected distances ($<$$50$~kpc) many of the extragalactic HVCs are not associated with regions of star formation suggesting that they too are infalling for the first time \citep{Thilker2004,Westmeier2008}.
Despite the abundance of other sources for HVCs, explaining why all dark matter subhaloes that failed to form stars lack any gas is difficult and a small fraction of HVCs may be such objects.

In order to investigate the dark matter content in \HI{} clouds, we recently searched for gamma-ray emission from dark matter annihilation at the location of several compact \HI{} clouds in the GALFA-HI Compact Cloud Catalogue \citep{Mirabal2013}.
However, due to poorly constrained dark matter profiles and unknown distances to these objects severely limited our analysis.
To advance more thoroughly, we need to examine systems with better distance determinations and are likely to be candidates for dark matter embedded HVCs.
The best candidate for such a search is the Smith Cloud, a massive \HI{} system near the Galactic disc \citep{Smith1963}.
The Smith Cloud is located close by at $12.4\pm1.3$~kpc \citep{Lockman2008}, and of particular appeal is that a dark-matter subhalo seems to be required for the survival of the gas cloud after a passage through the Galactic disc \citep{Nichols2009}.

Given its relative proximity, apparent orbit and large mass it appears to qualify as the ideal astrophysical site to test the dark matter confinement of \HI{} clouds.
Here we search for potential gamma-ray emission from the dark matter annihilation of weakly interacting particles (WIMPs) around the Smith cloud.
In Section 2 we present arguments in favour of a dark-matter subhalo surrounding the Smith 
cloud. Next, we present the {\it Fermi}-LAT analysis and 
derive {\it Fermi} upper limits to the gamma-ray flux. 
In Section 5, we turn the results into limits on annihilation 
cross sections. Finally, we summarise our results and 
present our conclusions.

\section{The Smith Cloud and a dark matter halo}
Most HVCs do not require dark matter to explain their survivability and formation, being able to form through galactic fountains, tidal stripping of gas or condensation of primordial gas.
We argue that none of these processes are likely to have created the Smith Cloud, and that a gas cloud encapsulated by dark matter provides the best model.

The Smith Cloud is an extraordinary gas structure with a cometary tail being accreted onto the Galaxy \citep{Lockman2008,Nichols2009}.
Unusually for a HVC, the distance to the Smith Cloud is relatively well know (to within $20\%$).
Based on stellar bracketing \citep{Wakker2008}, interaction of stripped material with disc gas \citep{Lockman2008} and the flux of H$\alpha$ reflected from the Galactic disc \citep{Bland-Hawthorn1998,Putman2003}, the distance to the Smith Cloud is known to be $12.4\pm1.3$~kpc \citep{Lockman2008}.

Here we use H{\small I} 21~cm data from a preliminary reduction of a new
survey of the Smith Cloud with the 100 meter Green Bank Telescope (GBT) of
the National Radio Astronomy Observatory \footnote{The National Radio
Astronomy Observatory is a facility of the National Science Foundation
operated by Associated Universities, Inc.}.   The new survey covers more
than twice the area of the \citet{Lockman2008} survey and uses a
different spectrometer that provides four times  the velocity resolution
($0.32$~km~s$^{-1}$) over a $1.6$ times greater LSR velocity range, $-400$ to
$+400$~km~s$^{-1}$~LSR.  The preliminary reduction of the new survey data has
a median noise level of 65~mK, a factor 1.5~times more sensitive than the
previous survey, and the observing and reduction techniques produce maps
at the full 9.1~arcmin resolution of the GBT.  The new survey is also
corrected for residual stray radiation \citep{Boothroyd2011} improving the dynamic range at low velocities.  The new
data show that the Smith Cloud is considerably more extended than
previously thought, with a diffuse tail to at least $l,b=50,-25$ and components that appear to have been detached from the main
structure (e.g. at $38,-18$ and $40,-22$).  The observations are nearing
completion and the full data set will be published separately.

The existence of a velocity gradient across the cloud allows a determination of the true $3D$ velocity (and consequent orbit) if the Smith Cloud is travelling along the plane of the sky.
The orbit is calculated using the potential from \citet{Wolfire1995} as detailed in \citet{Lockman2008}. 
Using the orbital determination of \citet{Lockman2008} the cloud will have passed through the disc approximately $70$~Myr ago at a radius of $13$~kpc.
Such an orbit assumes that the velocity gradient is towards the lower end, that drag is minimal along the Smith Clouds orbit, and the current distance estimate is correct.
We have examined the effects of changing the velocity gradient, distance, and introducing drag with coefficients of drag up to $C_D=5$.
Even with the most generous assumptions of drag, distance and velocity gradient, the total velocity of the Smith Cloud is restricted by observations to $\sim300$~km~s$^{-1}$ and so cannot have fallen in from infinity.
Furthermore, in all but the most extreme cases, the Smith Clouds orbit will pass through the disc at a radius of approximately $13$~kpc, and hence the cloud must have passed through the disc or been created in the past $\sim$$70$~Myr.

\subsection{Galactic fountains}

Compared to most of the \HI{} clouds that surround the Milky Way, the Smith Cloud is exceptionally massive, with a gas mass exceeding $2\times10^{6}~M_\odot$.
Correspondingly, the energy contained in the Smith Cloud is also extremely large compared to the gas at any potential launching point inside the Galactic disc ($\sim{}1.0\times10^{54}$~erg).
This tremendous energy requirement would require of order $1000$~supernova in order to produce sufficient energy to dislodge the Smith Cloud inside a galactic fountain.
Such a process would also have to take place at a radius $\sim$$13$~kpc (the distance where the Smith Cloud's orbit intersects the disc $\sim$$70$~Myr ago) in order to be produced \citep[such a simple calculation is in line with a linear scaling of galactic fountain hydrodynamic simulations by][]{Gouveia2009}.
The requirements of a massive cluster at a large radius hence makes it extremely unlikely that the Smith Cloud was formed through any galactic fountain.

That the Smith Cloud would be exceptionally rare if it was a product of the galactic fountain is further supported by the the fact that the largest observed \HI{} cloud elevated by a supershell is only $3\times10^4$~$M_\odot$ and, like most gas clouds associated with the galactic fountain, has a low peculiar velocity relative to the disc \citep{Pidopryhora2007}.

If it did not arise directly from the disc, it is still possible to have avoided a disc passage if it had undergone a tidal interaction in the recent past.
However, there is no obvious candidate to have provided such a massive cloud and given the paucity of dwarf galaxies that contain gas \citep{Grcevich2009}, the detection of a suitable host that until recently held gas would raise further questions.

\subsection{Passage through the disc}
Without arising from the disc and a tidal event being unlikely, the original origin of the Smith Cloud is likely to be from extragalactic infalling gas.
Such HVCs are potentially able to survive to near the disc without any dark matter through ablative shielding \citep{Putman2012,Plockinger2012}.
In particular, \citet{Plockinger2012} suggest that dark matter will suppress the characteristic substructure present within HVCs as they approach the disc.
However, little work has been done on the passage of HVCs through the disc, potentially the biggest impediment to the survival of the Smith Cloud.

We note that the current cloud total velocity relative to the expected Milky Way virial velocity $|V_{\rm tot}/V_{\rm vir}|\sim2$, compatible with phase space data of substructure at $r\sim 10$ kpc in simulated high resolution dark matter haloes of Milky Way mass \citep{Boylan-Kolchin2012}.

At $13$~kpc, the surface density of the disc is approximately $3\times10^{20}$~cm$^{-2}$ \citep{Kalberla2008} and is increased by a factor of three due to the oblique angle at which the Smith Cloud passes through the disc.
At such high column densities gas clouds passing through the disc are likely to be heavily disrupted.
Simple toy models presented in \citet{Nichols2009} suggest that this disc passage and ram pressure stripping throughout the orbit will require a dark matter halo to survive to the present day.

We supplement the toy model argument presented previously, which suggested a dark matter halo of $M_{\rm tid}\sim2\times10^{8}$--$1\times10^9$~$M_\odot$, with numerical simulations of a cloud passing through a disc with and without a supporting dark matter halo.

\subsection{Numerical simulations}
\begin{figure}
\includegraphics[width=0.475\textwidth]{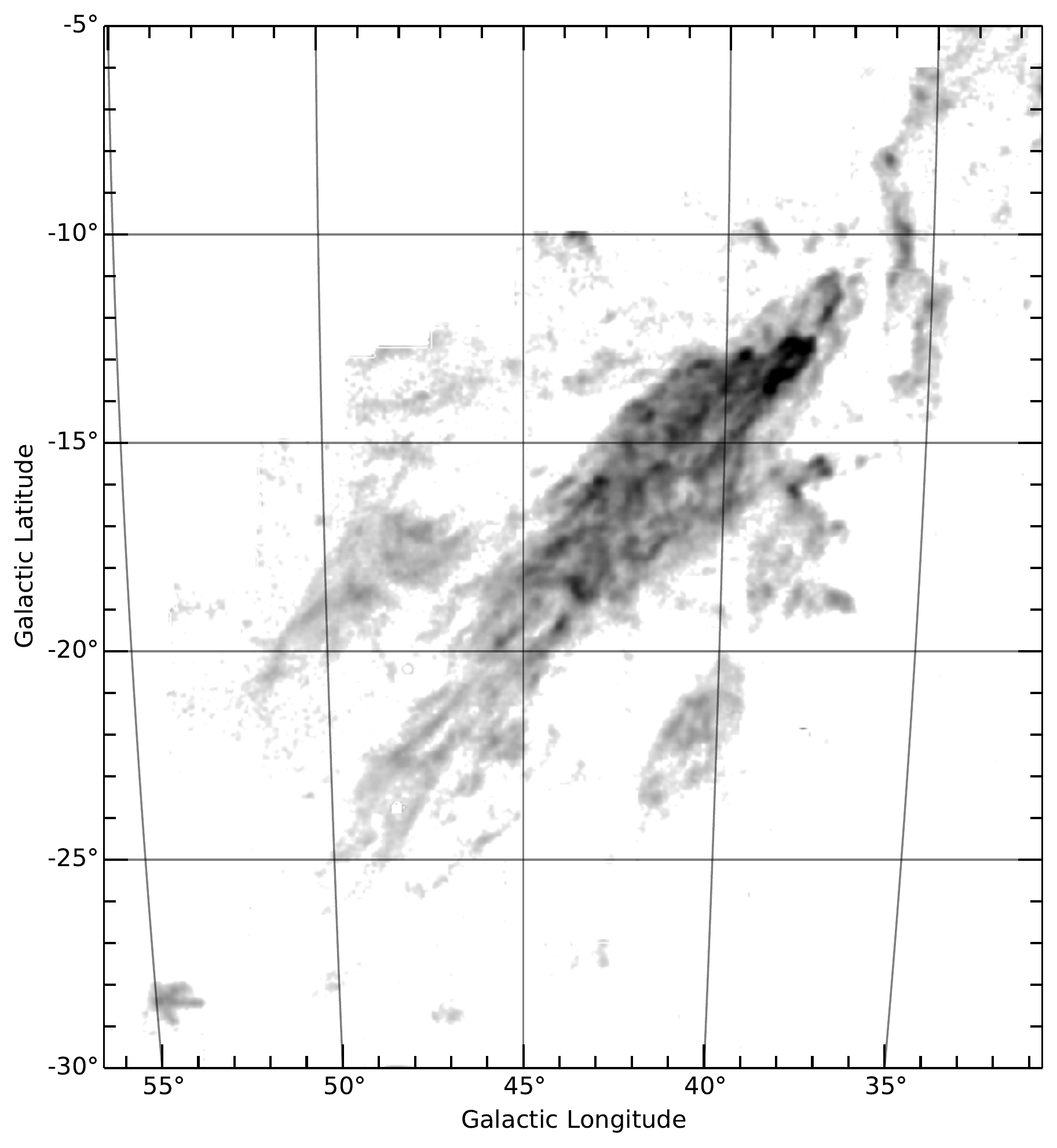}	
\caption{GBT H{\scriptsize I} image of the Smith Cloud integrating over all $V$$_{\rm GSR} > 220$~km~s$^{-1}$. The grey scale is proportional to the square root of the H{\scriptsize I} column density to reveal the extended cometary morphology.}
\label{fig:smithraw}
\end{figure}

In an attempt to replicate the observed mass, morphology and \HI{} distribution of the Smith Cloud today, shown in Fig. \ref{fig:smithraw}, we use the Adaptive Mesh Refinement (AMR) code {\tt RAMSES} \citep{teyssier2002}.

Using {\tt RAMSES} we simulate the disc passage of both a dark matter enclosed \HI{} cloud and a dark matter free cloud, along the Smith Cloud's expected orbital path from peak height above the disc $204$~Myr ago and compare that to new \HI{} observations.
We include a realistic progenitor to the Smith Cloud and a realistic model of the Milky Way which includes live stellar and gaseous discs and dark matter halo. We evolve the coupled $N$-body and hydrodynamical system while accounting for metal dependent radiative cooling and UV background heating as per \cite{Agertz2013} at a maximum resolution of $\Delta x \sim18$~pc.

The Smith Cloud progenitor is modelled over a range of densities, $n=0.1$--$0.5$~cm$^{-3}$ in $0.1$~cm$^{-3}$ steps, with a constant size, $r=0.5$~kpc, in order to explore the potential parameter space of the Smith Cloud.
Such a range ($M_{\rm HI,initial}=1.3-7.8\times 10^6\,M_\odot$) includes clouds approximately half the mass of the Smith Cloud and to clouds approximately three times bigger than the Smith Cloud today, but in line with massive clouds in nearby systems \citep[\eg][]{Chynoweth2008} as well as allowing a large degree of mass loss to the present day.
In order to model the survival with and without a dark matter halo, we consider both a pressure supported cloud and one which is embedded within a Navarro-Frenk-White (NFW) dark matter halo of virial mass $M_{\rm NFW}=3\times10^{8}$~$M_\odot$ and concentration $c=30$.

The model galaxy is considered to consist of a galactic disc of total mass $M_{\rm disc}=4.3\times 10^{10}~M_\odot$, where the gaseous disc accounts for $8.6\times 10^9\,M_\odot$ of the baryons. The disc has a scale radius of $r_{\rm disc}=3.4$~kpc and an exponential scale height $h=0.1r_{\rm disc}$.
These discs are embedded inside a hot gaseous halo with density $n=10^{-3}$~cm$^{-3}$, a reasonable density out to $r\sim 10-30$ kpc \citep{Gatto2013}, along with a $M_{\rm vir}=10^{12}$~$M_\odot$ NFW dark matter halo with concentration $c=10$. Note that the adopted gaseous halo density, out to the initial simulation cloud altitude ($z\sim 20$~kpc) 204 Myr ago, is a \emph{conservative} choice, as the observed circum-disc medium likely is denser; \cite{Gaensler2008} found that a warm gaseous disc of the form $n(z)=n_0\exp(-z/H)$, where $n_0=0.014$~cm$^{-3}$ and $H=1.8$~kpc, is consistent with the Milky Way. 

\begin{figure}
  \includegraphics[width=0.5\textwidth]{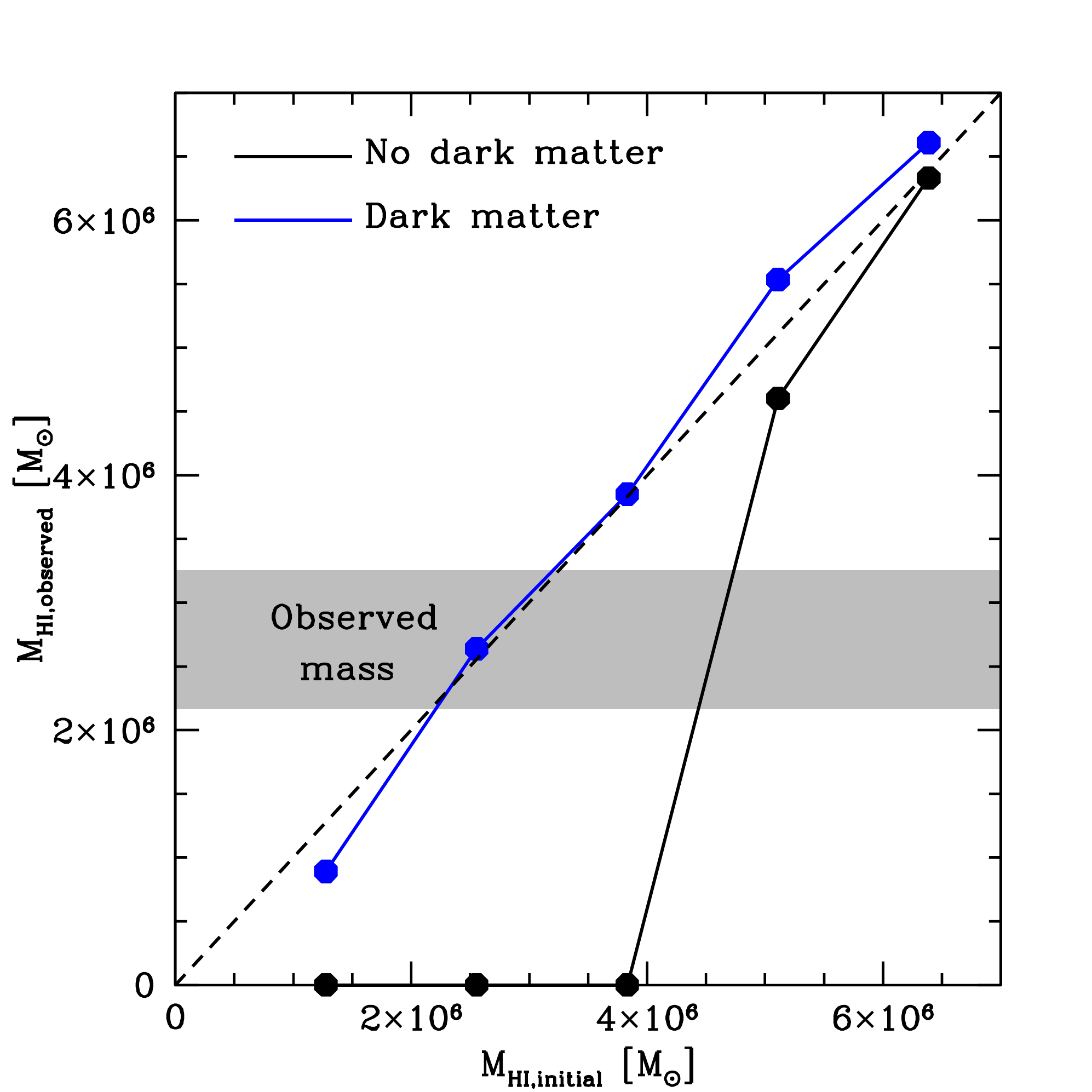}
  \caption{Initial H{\scriptsize I} mass versus the mass that would be observed today.
Dark matter free models are shown as black points (lower line) and dark matter models in blue (upper line).
The dashed line shows a 1:1 correspondence, points below this represent clouds which have lost mass, and points above will have accreted mass over the simulation.
The observed mass of the Smith Cloud is shown as a grey bar as calculated from the H{\scriptsize I} distribution of the Smith Cloud.}
  \label{fig:final-mass}
\end{figure}

In Fig. \ref{fig:final-mass} we display the initial and the \HI{} mass that would be observed today of the simulated clouds.
Dark matter free clouds with densities below $n=0.4$~cm$^{-3}$, initial masses below $5\times10^{6}$~$M_\odot$, are entirely destroyed by their passage through the hot halo and Galactic disc.
Clouds with densities above this, survive the passage, with nearly all of their mass intact.
In contrast to the baryon-only clouds, the dark matter supported clouds at all masses manage to pass through the disc of the Galaxy with minimal mass loss, or at higher ends, a small gain due to gas accreted from the disc.

\begin{figure*}
\hskip2.3cm\textbfss{n=0.2~cm}\mathbfss{$^{-3}$}\hskip4.325cm\textbfss{n=0.3~cm}\mathbfss{$^{-3}$}\hskip4.325cm\textbfss{n=0.4~cm}\mathbfss{$^{-3}$}\hfill{}\phantom{.}
\\
\includegraphics[width=0.33\textwidth]{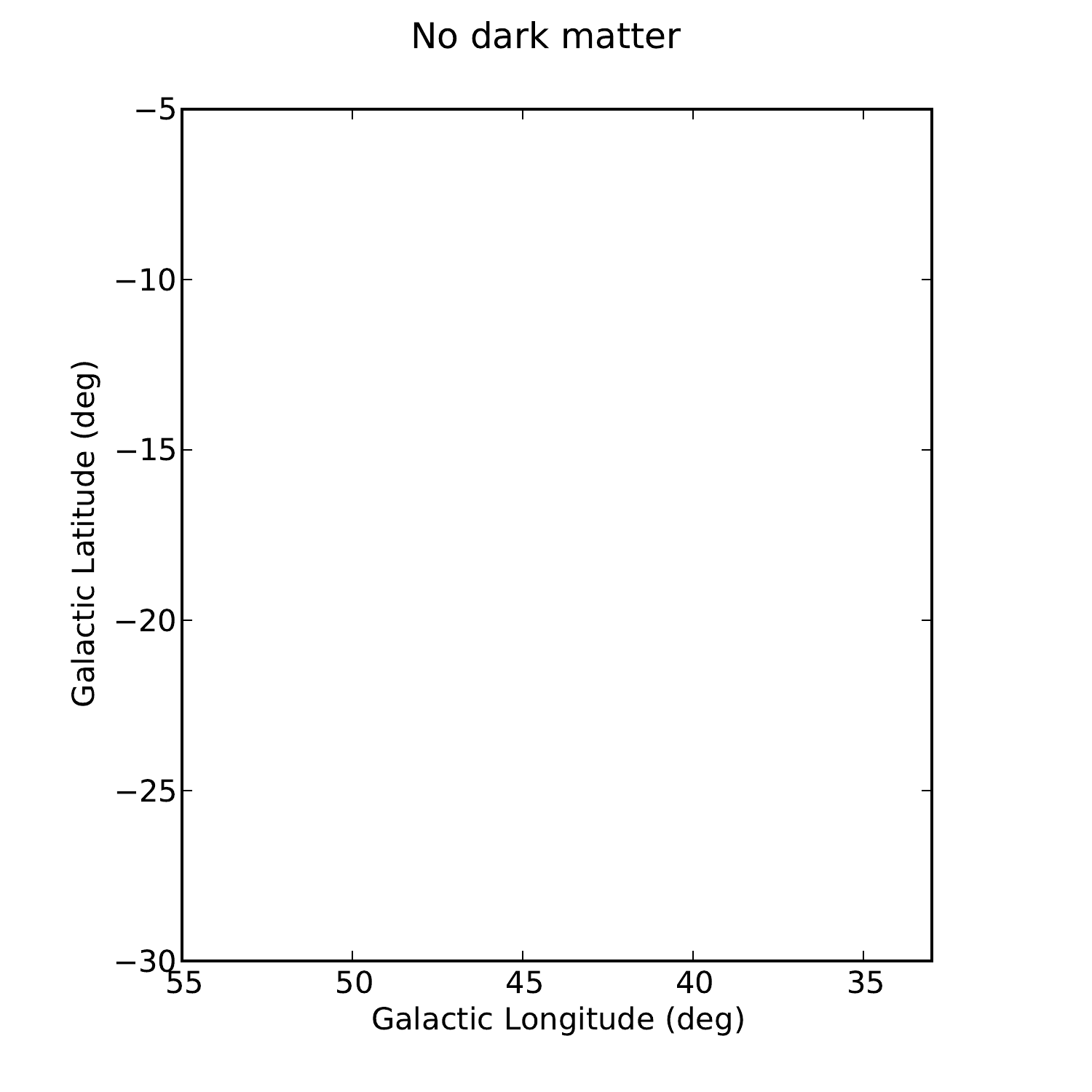}
\includegraphics[width=0.33\textwidth]{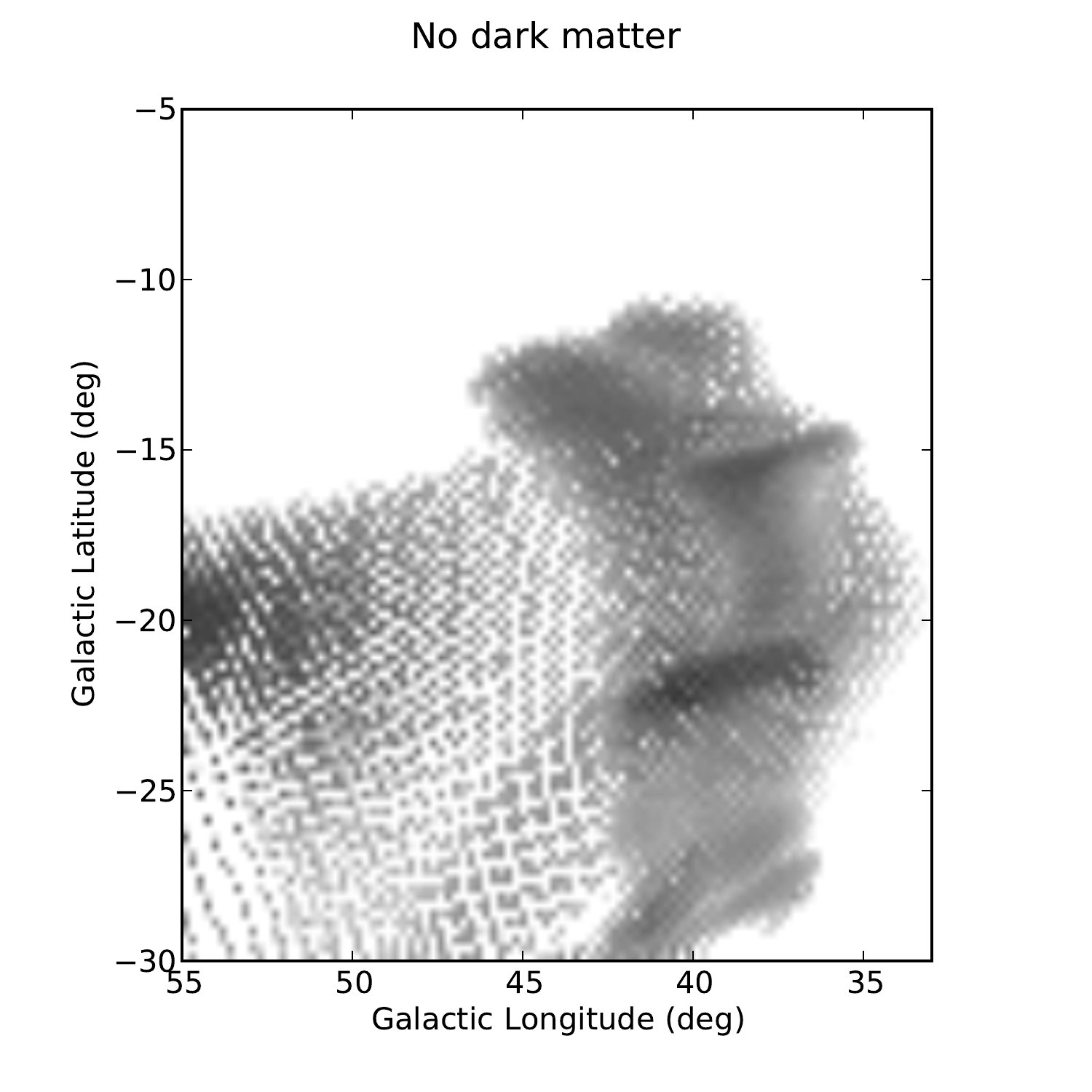}
\includegraphics[width=0.33\textwidth]{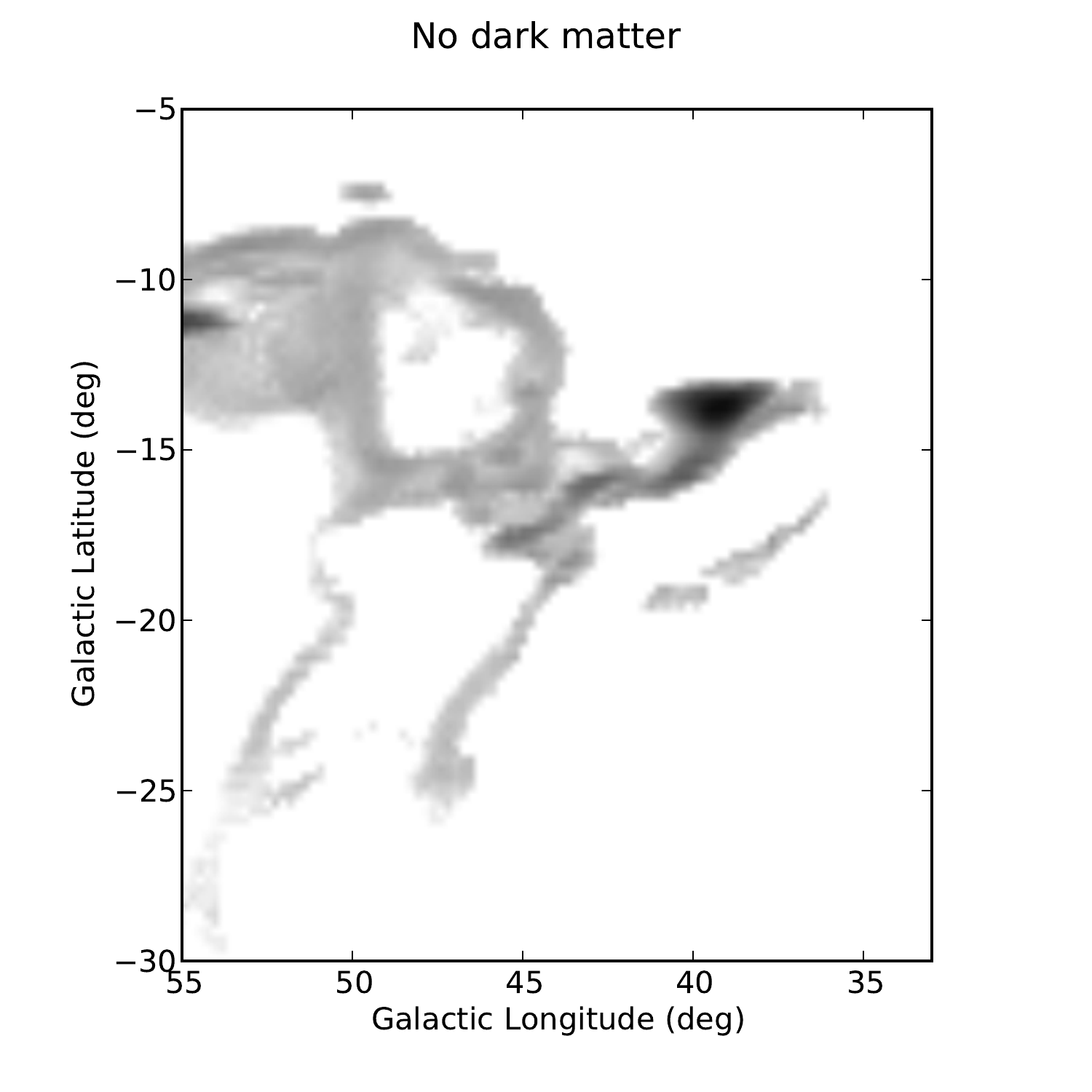}
\\
\includegraphics[width=0.33\textwidth]{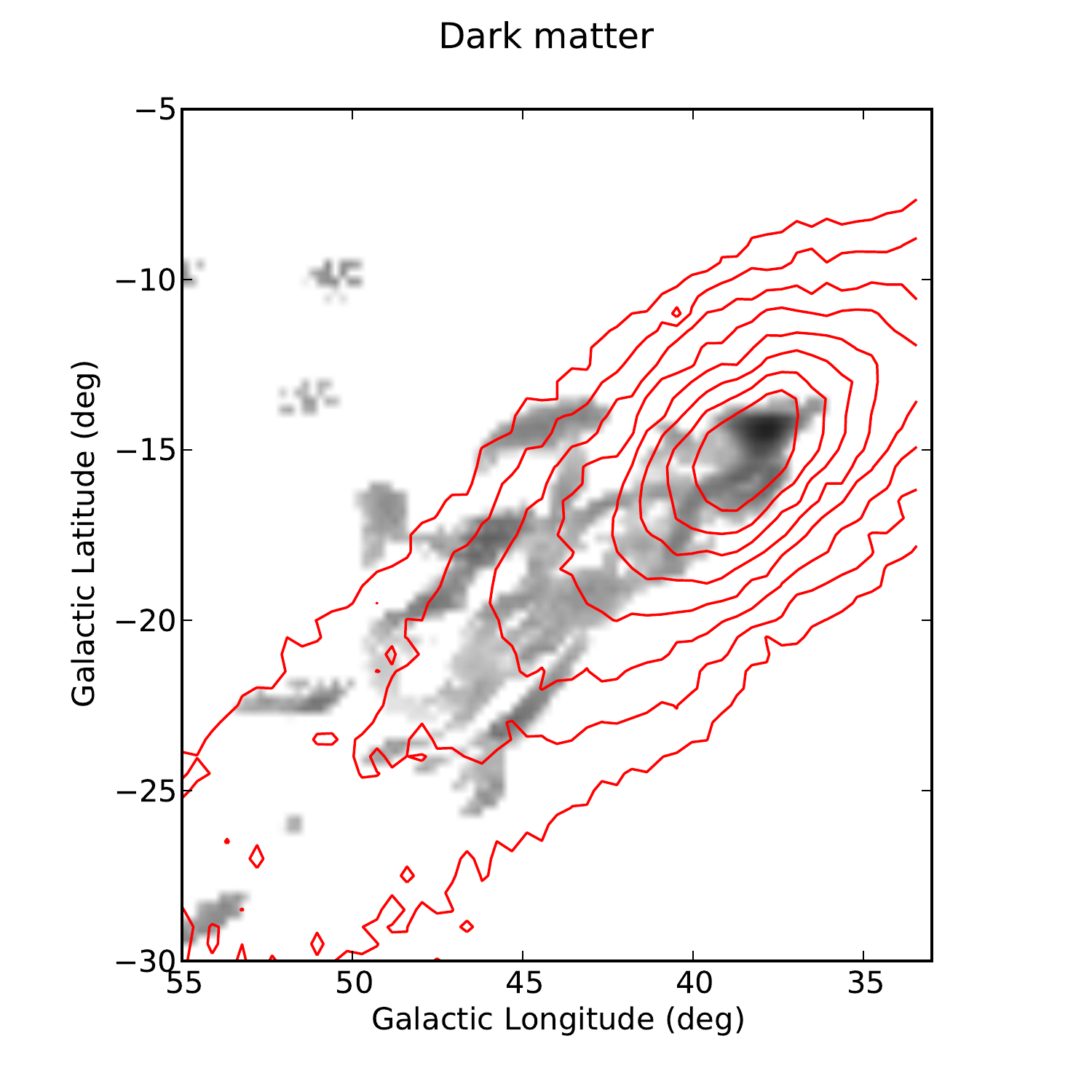}
\includegraphics[width=0.33\textwidth]{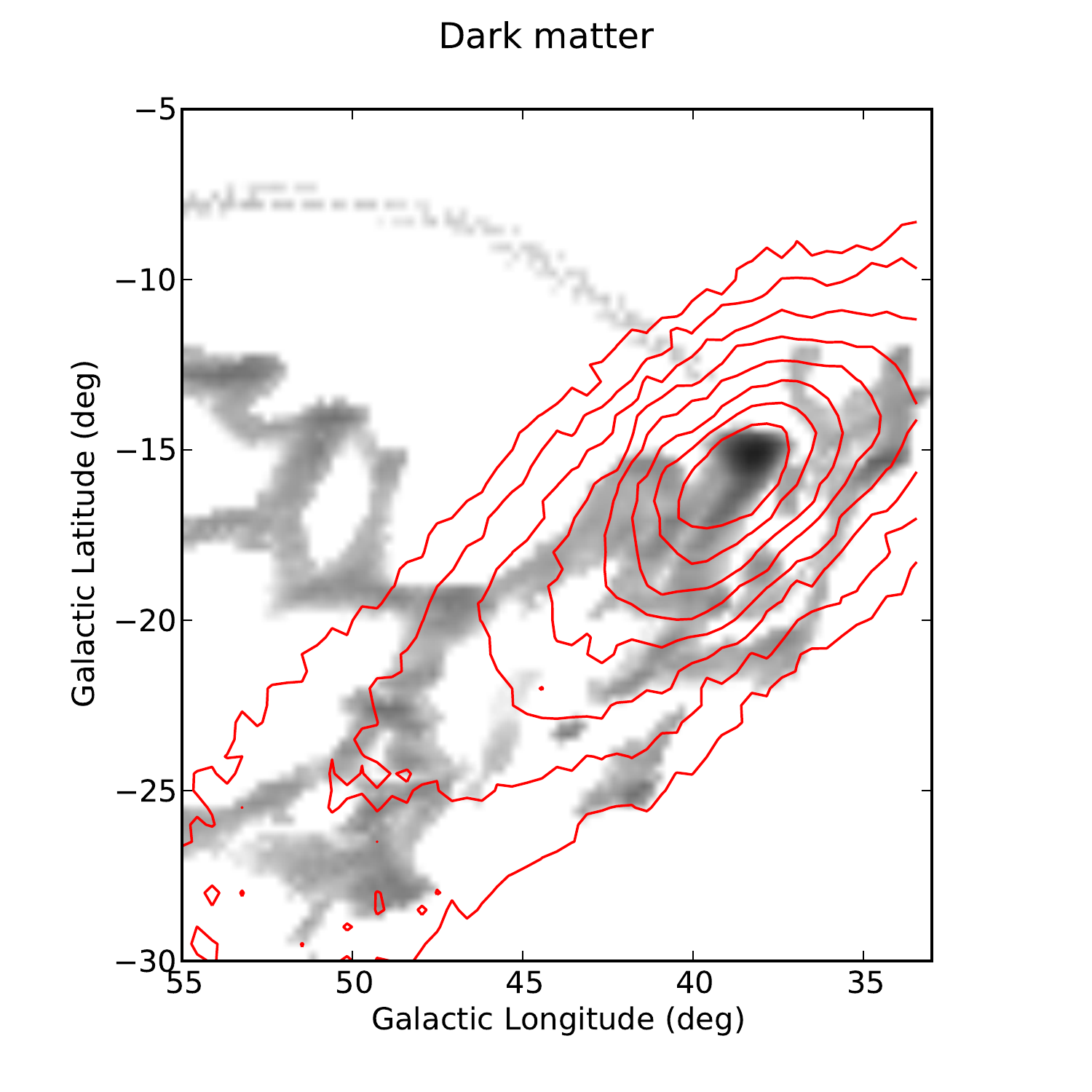}
\includegraphics[width=0.33\textwidth]{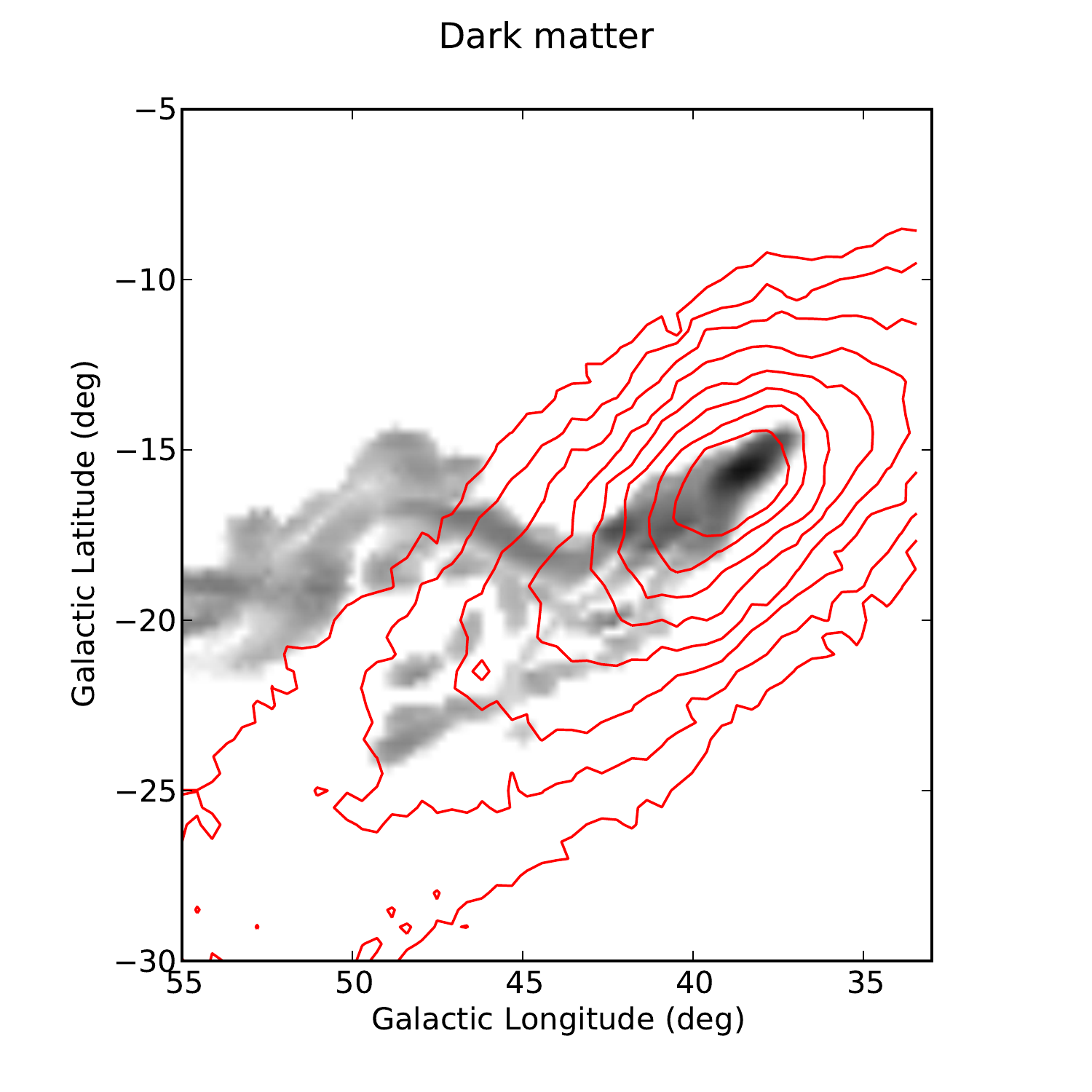}
\caption{Simulated H{\scriptsize I} intensity for a HVC that has passed through the Galactic disc, without dark matter (top row) or encapsulated by a dark matter halo (bottom row).
The columns correspond to the initial density of the H{\scriptsize I} cloud, from left to right, $n=0.2$~cm$^{-3}$, $n=0.3$~cm$^{-3}$, and $n=0.4$~cm$^{-3}$.
Each plot is scaled to the simulation maximum H{\scriptsize I} column density.
In the dark matter present cases, the contours, shown in (red) solid lines, represent the average line of sight dark matter density between $10^8$~$M_\odot$~kpc$^{-3}$ and $10^{6}$~$M_\odot$~kpc$^{-3}$ in $0.25$~dex steps.}
\label{fig:morphology} 
\end{figure*}

The morphological features of the \HI{} are also examined.
In Fig. \ref{fig:morphology} we show the observed \HI{} density today (rescaled to maximum density in each case), for both dark matter free clouds (top row) and dark matter encapsulated clouds (bottom row). We display only the middle three densities as these adequately show the evolution of the cloud with mass.
 
At the lowest density considered, $n=0.1$~cm$^{-3}$ (not shown), the dark matter free cloud is destroyed, while in the dark matter case, the \HI{} is completely separated from the dark matter halo and lacks the dense features reminiscent of the Smith Cloud. Such a stripped cloud will then likely be destroyed at the next disc passage as a dark matter free cloud.

At $n=0.2$~cm$^{-3}$, the dark matter free cloud is again destroyed, with no observable remnant surviving. In contrast, the dark matter supported cloud survives, with a dense core and recently shed cloudlets, producing an extended fragmented structure reminiscent of the observed cloud.

At $n=0.3$~cm$^{-3}$, the dark matter free cloud retains little mass, but has an observable structure. However, this cloud retains only a smooth, low column density structure. The dark matter encapsulated cloud meanwhile retains the dense clumps but, in addition to being more massive than the observed cloud, suffers little fragmentation, with long tendrils that would be readily observable.

At $n=0.4$~cm$^{-3}$, the dark matter free cloud is massive enough to survive its passage through the disc of the Galaxy, and begins to fragment. While portions of it follow the orbital path, a large portion of it has been stripped, and suffering from drag extends the cloud to higher Galactic Longitudes at approximately constant latitude. Such a structure is not seen in the Smith Cloud today, and once again contains too much mass. The dark matter cloud again displays the tendrils with little fragmentation compared to the central clump, but again contains too much mass to be equivalent today.

At $n=0.5$~cm$^{-3}$ (not shown), the clouds are almost identical. Both display the tendrils associated with high mass clouds, as well as a dense core. In addition to the morphological differences, both clouds contain too much mass to be representative of the Smith Cloud today.

\begin{figure*}
  \includegraphics[width=0.425\textwidth]{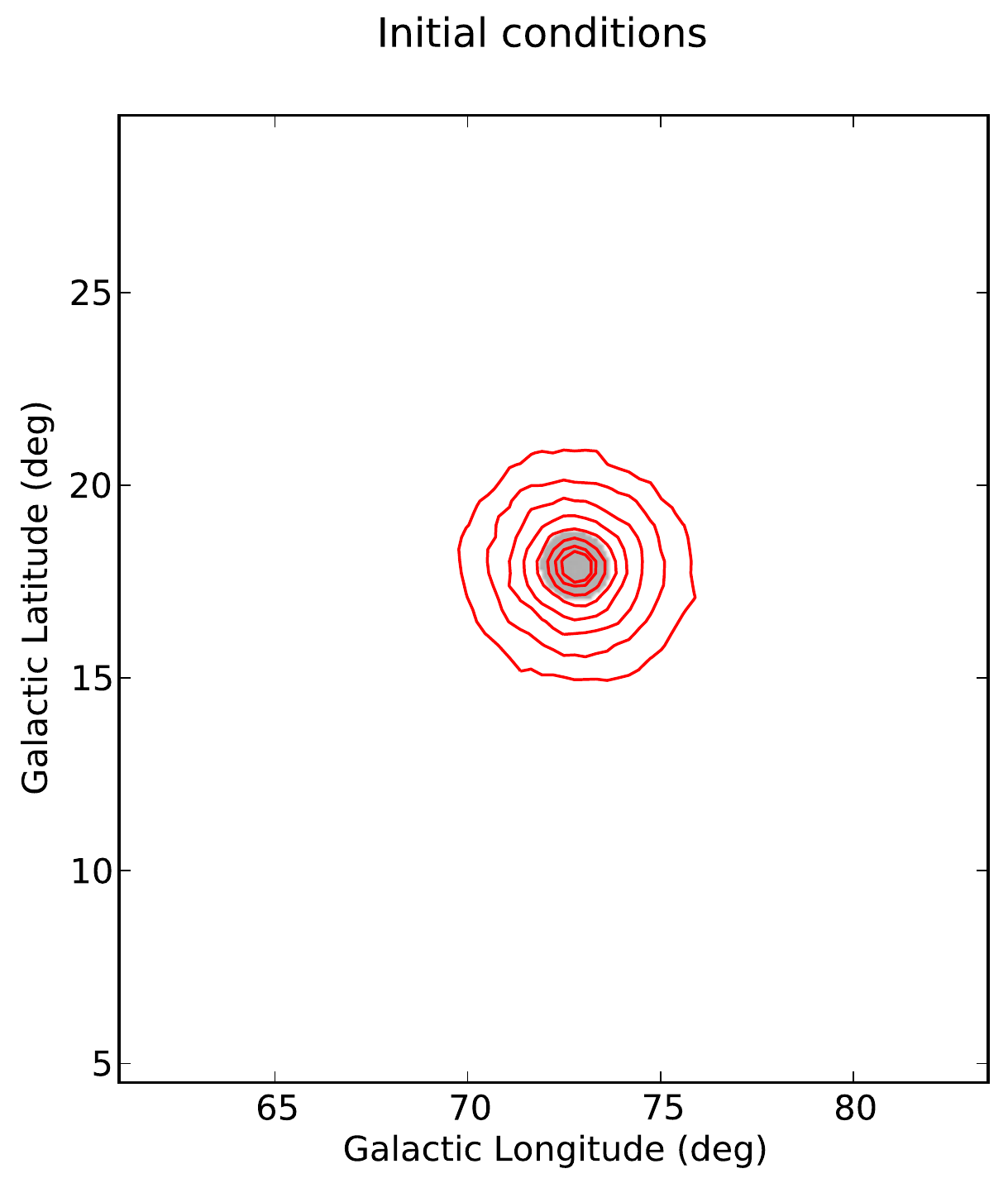}\hfill
  \includegraphics[width=0.425\textwidth]{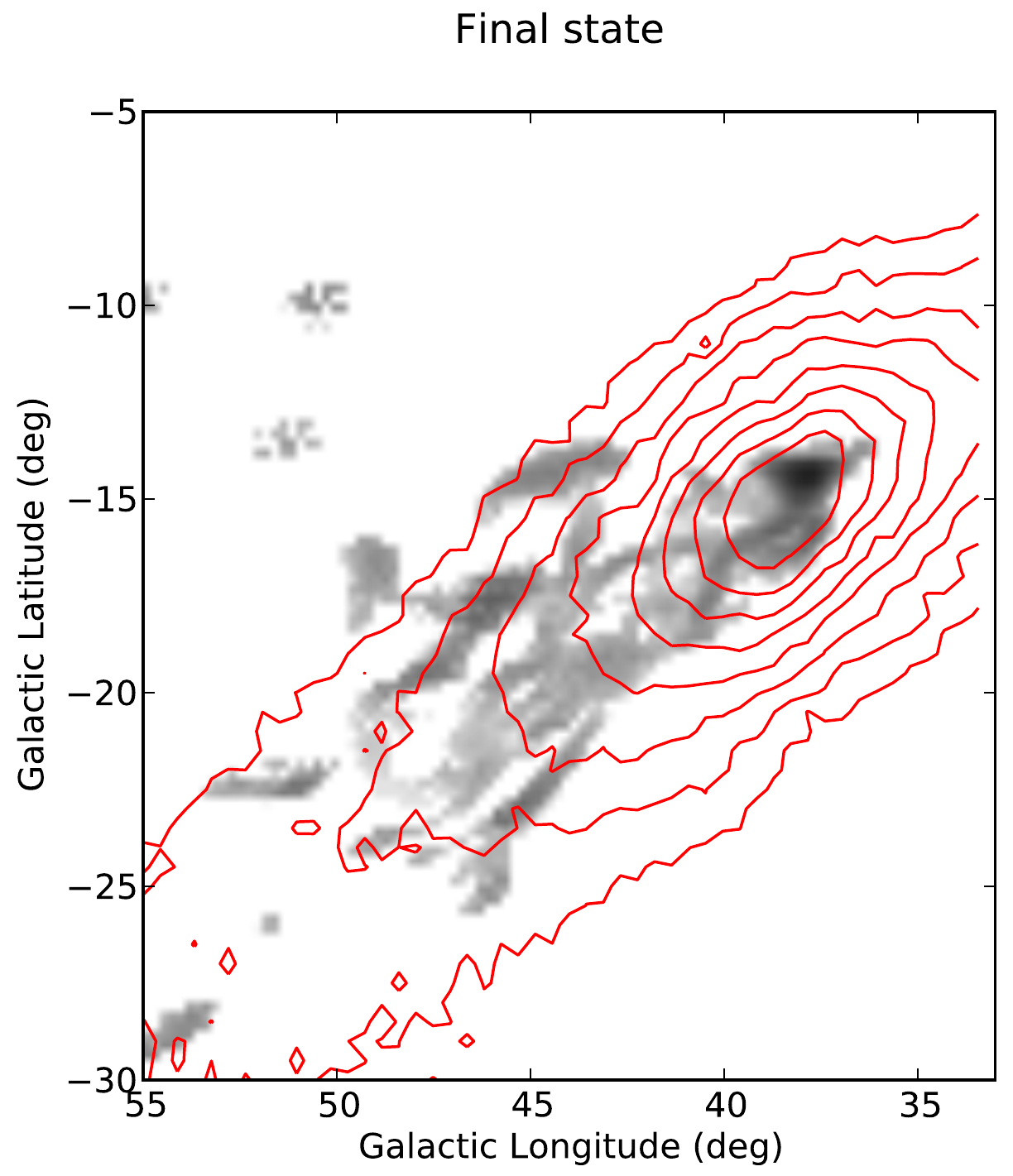}
  \caption{Initial and final state (for the best fit $n=0.2$~cm$^{-3}$ model) for the simulated cloud, with dark matter contours as in Fig. \ref{fig:morphology}.
The greyscale represents the relative column density across both clouds and is not equivalent between the figures.
The initial conditions are the same for all models with only the gas density changed as required.
In addition to the gas being disturbed by ram pressure and passage through the disc, the dark matter is also disturbed due to the gravitational forces on the disc.}
\label{fig:DMdistortion}
\end{figure*}

Due to the added gravitational attraction of the baryonic disc the dark matter core of the HVC becomes significantly elongated.
We show this distortion in Fig. \ref{fig:DMdistortion}, comparing the initial conditions with the final result for a dark-matter encapsulated cloud with a density of $n=0.2$~cm$^{-3}$.
The dark-matter subhalo finishes with a flattening of $\sim$$1/2$ due to tidal forces of the disc.
Such a transformation from a spherical halo to a roughly prolate spheroid differs from the expected result from dark matter only simulations, where tidal forces tend to make a subhalo more spherical \citep{Moore2004}.

\begin{figure*}
\includegraphics[width=0.45\textwidth]{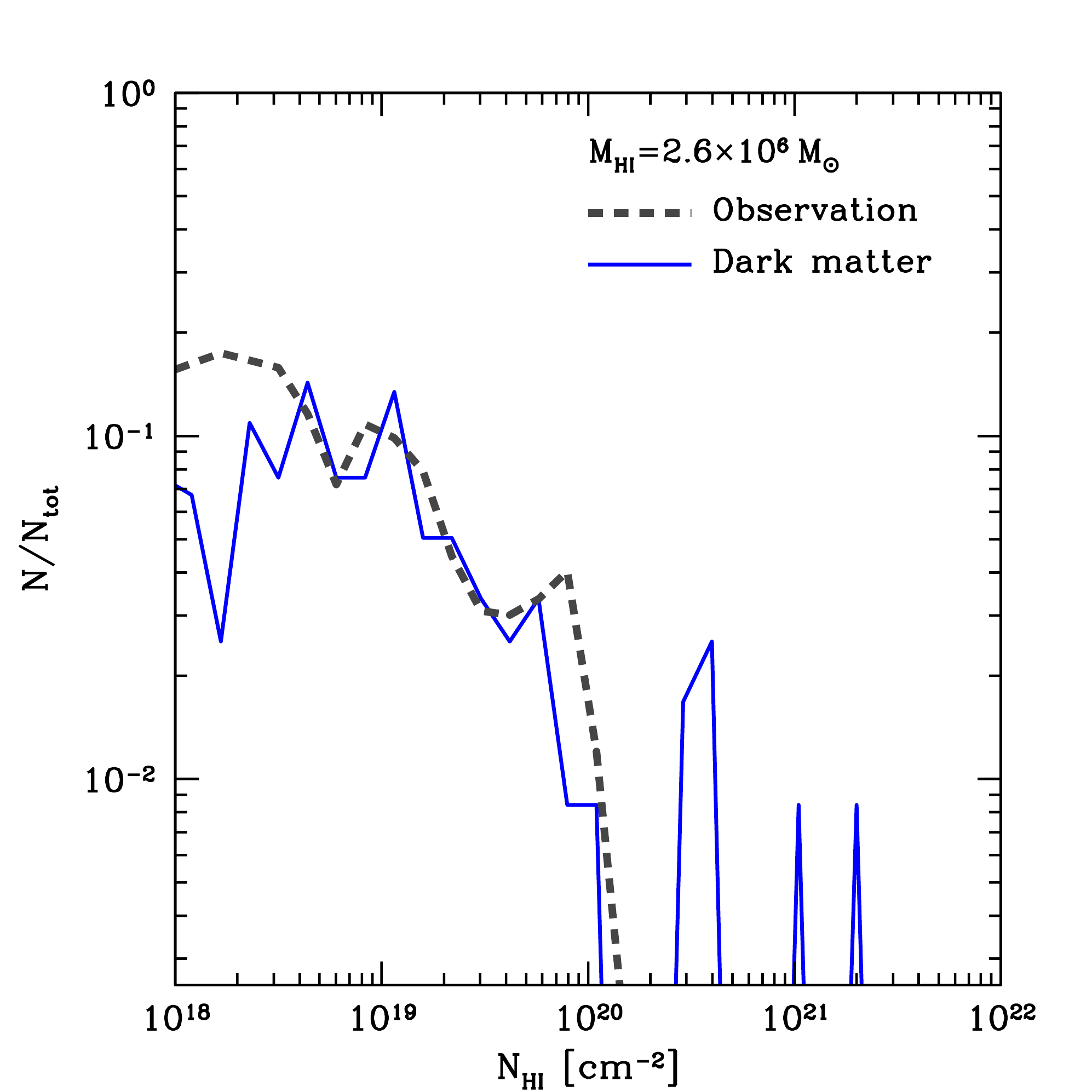}\hfill
\includegraphics[width=0.45\textwidth]{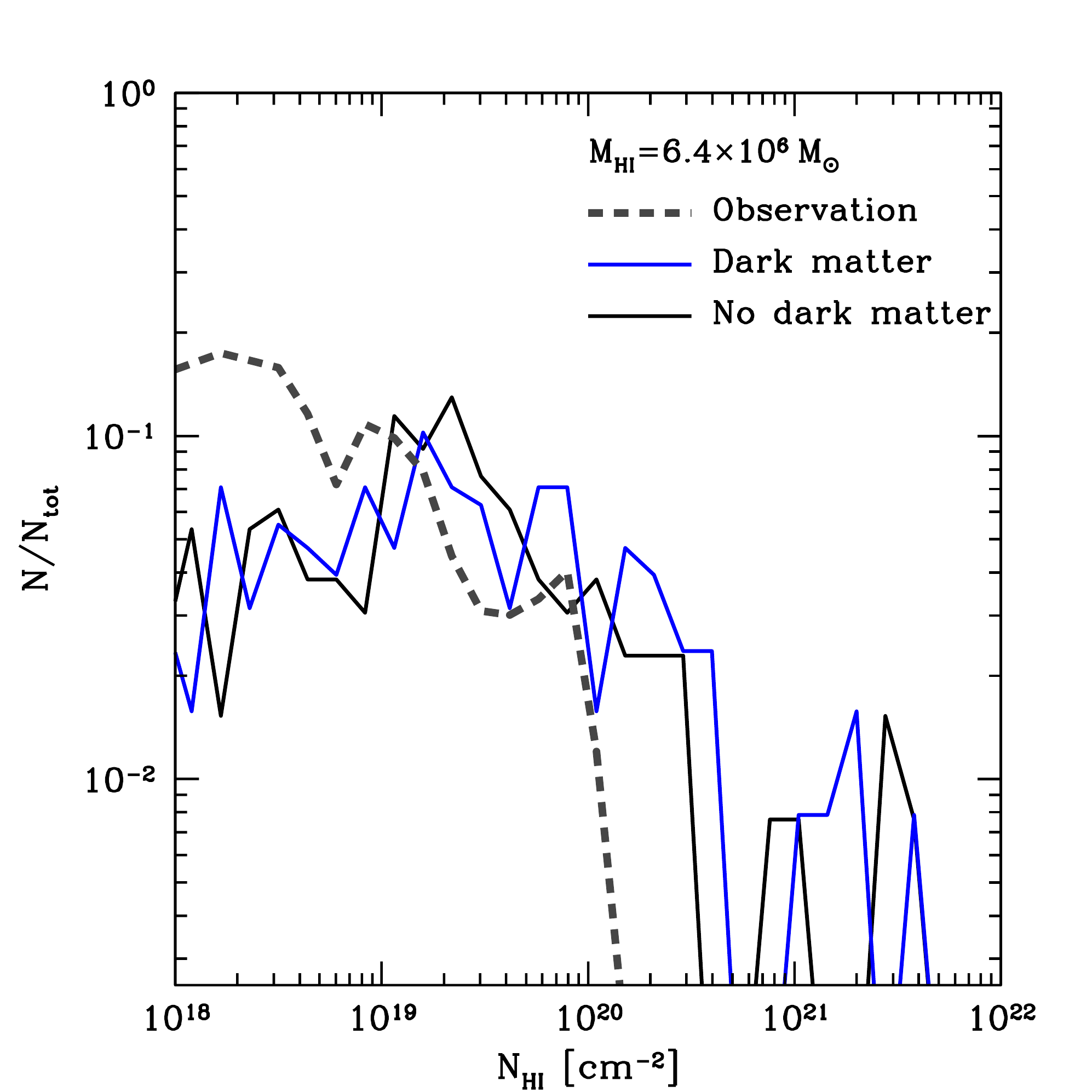}
\caption{Mass weighted probability function of H column density within the modelled clouds.
The left panel shows a $n=0.2$~cm$^{-3}$, $M_{\rm HI}=2.6\times10^6$~$M_\odot$, dark matter embedded cloud in blue.
The right panel shows $n=0.5$~cm$^{-3}$, $M_{\rm HI}=6.4\times10^6$~$M_\odot$, dark matter free in black and dark matter embedded clouds in blue.
In both panels the observed distribution of the Smith Cloud is shown as a grey dashed line.}
\label{fig:gaspdf}
\end{figure*}

The column density weighted probability density function is shown in Fig. \ref{fig:gaspdf} for both the best fit, dark matter embedded model (left panel; $n=0.2$~cm$^{-3}$, $M_{\rm HI,obs}=2.6\times10^6$~$M_\odot$), and the high density model (right panel; $n=0.5$~cm$^{-3}$, $M_{\rm HI, obs}=6.4\times10^6$~$M_\odot$) both with and without dark matter.
The low density, dark matter embedded model follows the observed distribution remarkably well, replicating the general shape and drop off at column densities exceeding $1\times10^{20}$~cm$^{-2}$, with only occasional clumps above this limit.
Compared to this, both high density models don't begin to drop off until $\sim3\times10^{20}$~cm$^{-2}$, and retain clumps of gas above this limit.
Such clumps approach or exceed the lower limit of star formation free regions in dwarf galaxies \citep{Ekta2008} suggesting that such a cloud would begin forming stars regardless of the dark matter content.

\begin{figure}                                                                
\includegraphics[width=0.45\textwidth]{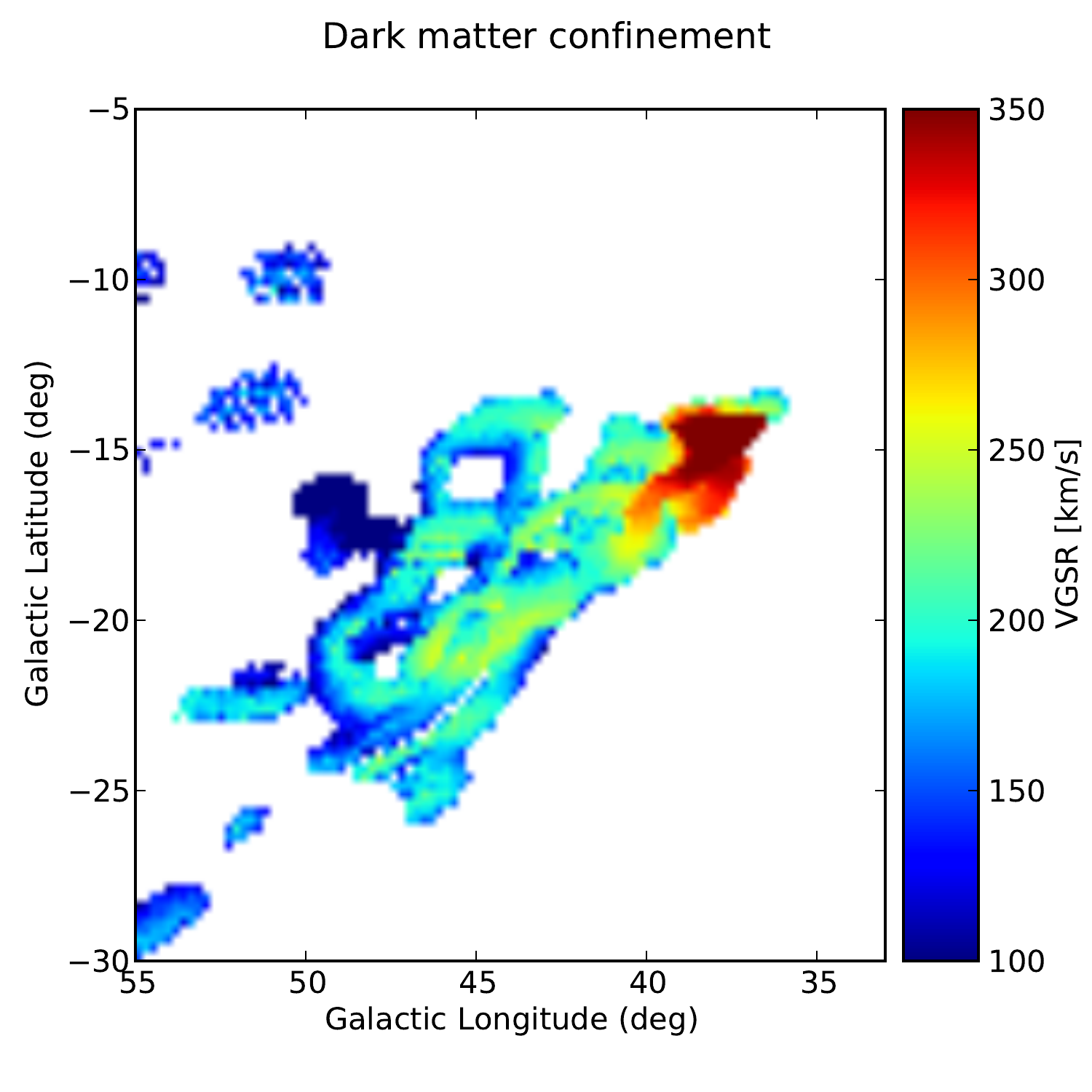}
\caption{H{\scriptsize I} $V_{\rm GSR}$ for a $2.6\times10^6$~$M_\odot$ HVC that has passed through the Galactic disc, encapsulated by dark matter halo. This model was the best fit of the simulation runs, showing morphological structure reminiscent of the Smith Cloud. The model cloud encapsulated by dark matter has a velocity gradient as observed in the present day Smith Cloud \citep{Lockman2008}, although the model gradient is slightly steeper.
}
\label{fig:vgsr}
\end{figure}

Shown in Fig. \ref{fig:vgsr}, the best fit model also displays a velocity gradient reminiscent of the observed cloud.
The velocity gradient is shown in terms of $V_{\rm GSR} = V_{\rm LSR} + 220\sin(l)\cos(b)$~km~s$^{-1}$, with the simulated cloud having a velocity gradient $\sim$7~km~s$^{-1}$ per degree.
This is in contrast to the observed Smith Cloud which possess a velocity gradient $\sim$4~km~s$^{-1}$ per degree \citep[see Fig. 3][]{Lockman2008}.
However, due to the large degree of variation in both figures, such a discrepancy is not significant. 

\subsection{Dark matter properties}
Given the above problems for a dark matter free Smith Cloud, the assumption of an encapsulating dark matter halo remains a viable hypothesis.
A simple range on the mass of the dark matter halo can be calculated by examining two boundary conditions.
If the halo is not massive enough it will face the same problems as a dark matter free cloud and will quickly smooth out after passing through the disc.
If it is too massive, the potential well will have enabled star formation to take place inside the core of the Smith Cloud, something for which there is no evidence.

Analytic limits are explored in \citet{Nichols2009} which, for an NFW profile, gives limits corresponding to a dwarf galaxy sized dark matter halo $M_{\rm vir}=10^9~M_\odot$, although this may vary by at least a factor of $3$. Such a halo corresponds to $r_s\sim1$--$1.8$~kpc and $\rho_s\sim6\times10^{6}$--$2\times10^{7}$~$M_\odot$~kpc$^{-3}$.

\begin{figure*}                                                                \

\hfil
\includegraphics[width=0.45\textwidth,angle=0.]{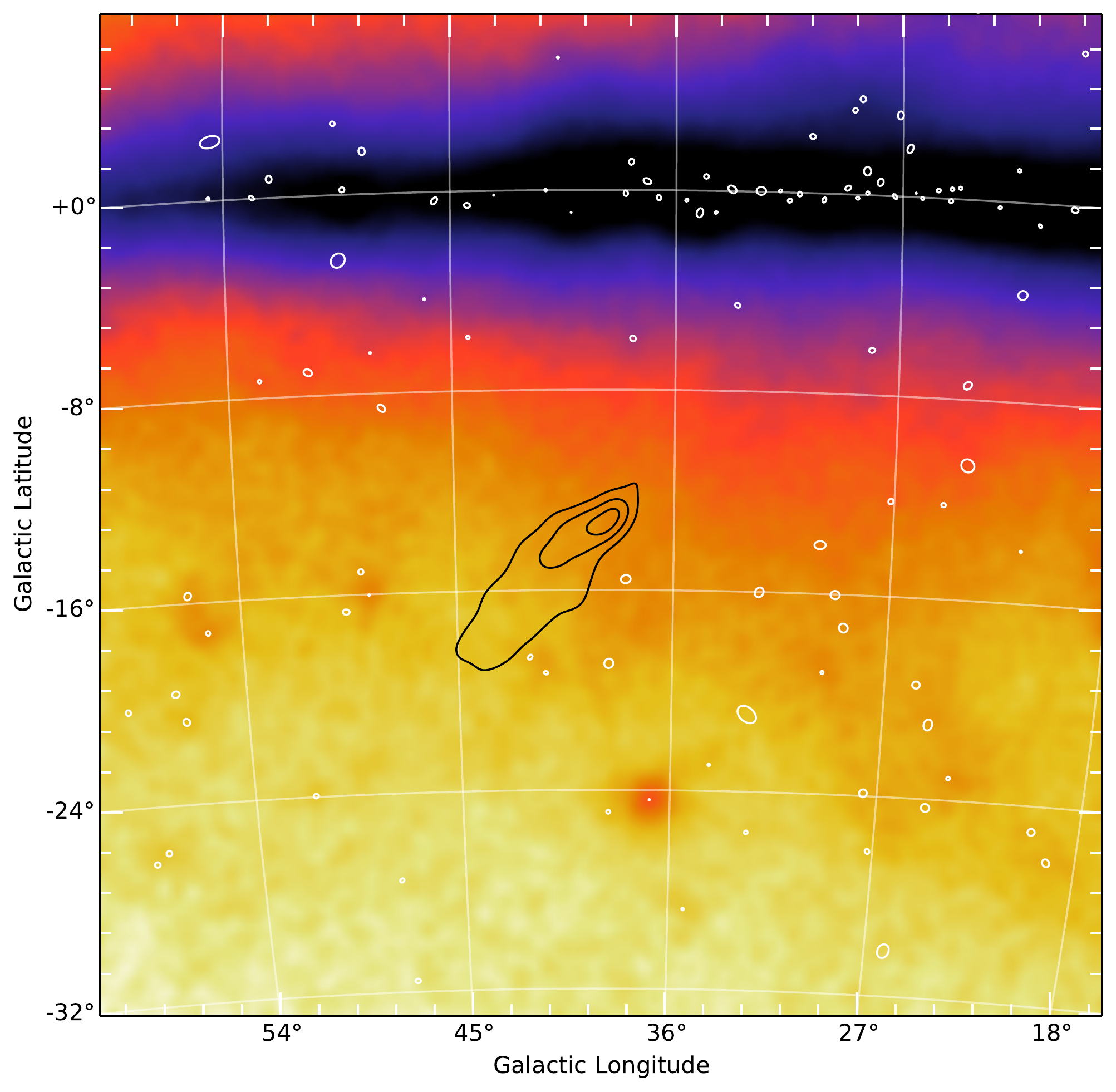}\hfill
\includegraphics[width=0.45\textwidth,angle=0.]{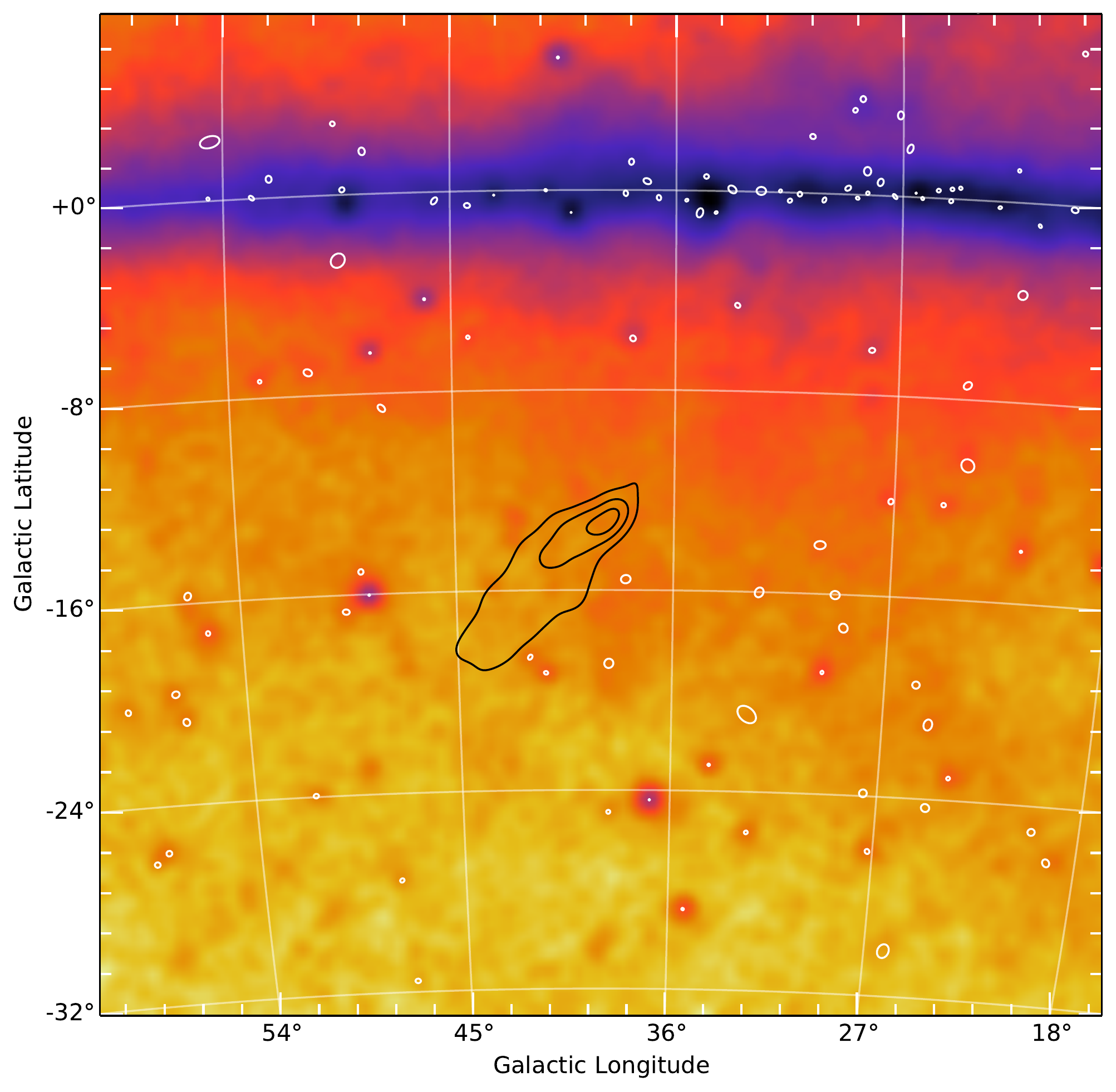}
\hfil
\caption{Smoothed {\it Fermi}-LAT count maps of a $40^{\circ} \times 
40^{\circ}$ region 
around the simulated Smith Cloud for 
100 MeV--1 GeV ({\it left}) and 1--300 GeV ({\it right}) respectively.
H{\scriptsize I} contours for the Smith Cloud are overlaid in green.
As the dark matter halo closely follows the H{\scriptsize I} distribution, this is also approximately the expected dark matter location. The white
symbols pinpoint 95\%
confidence ellipses of 2FGL point sources \citep{2lat}.
}
\label{figure1}
\end{figure*}

\section{Fermi-LAT observations and analysis}
\label{fermi}
To search for gamma-ray emission associated with the Smith cloud, 
we use {\it Fermi}-LAT observations carried out between 
2008 August 4 to 2013 June 11 (4.85 years of data), 
corresponding to the interval
from 239557417 and 392621346 in Mission Elapsed Time (MET). 
The analysis is performed using the \texttt{v9r27p1} {\it Fermi} Science Tools
together with  
the standard
\texttt{P7SOURCE\_V6} instrument response function.
In order to minimize contamination by the Earth's limb we  
exclude events with zenith angles $>$ 100$^{\circ}$.
We also filter the data using the \texttt{gtmktime} filter expression
recommended by the LAT team, namely ``{(DATA\_QUAL==1) \&\& (LAT\_CONFIG==1)
\&\& ABS(ROCK\_ANGLE)$<$52}''.
Fig. \ref{figure1} shows photon count maps for 100~MeV -- 1~GeV 
and 1 -- 300 GeV with contours of integrated H{\small I} emission from the Smith Cloud superimposed in green. With the current {\it Fermi} data, 
we do not see any obvious 
morphological correspondence between the gamma-ray emission and
the observed HI cometary 
morphology.

An interesting possibility is that there may be additional
spatially extended emission if the cloud was removed from the dark matter due to the shock of passing through the disc. In order to constrain such scenario, 
we sum gamma-ray counts over the 
entire orbit of the Smith cloud.
We consider photons within $\pm 1^{\circ}$
along the orbit, roughly corresponding to the 95\% containment angle
of the {\it Fermi} LAT 
for energies above 10 GeV. Point and and extended 2FGL sources listed
in \citet{2lat} are 
masked since they can contaminate the actual cloud contribution. The
counts are then compared with the observed counts in 
the anti-orbit of the cloud, corresponding to the same Galactic 
longitude but with opposite sign in Galactic latitude.
Compared
to its anti-orbit, we find no statistically 
significant gamma-ray excess in the orbit for energies above 1 GeV.
The presence of gamma-ray point sources present along the 
vicinity of the orbit
masks important portions of the trajectory, but at present
this seems the only reasonable procedure.

Given the absence of extended gamma-ray emission, events are only extracted
within a $10^{\circ}$ radius centred on the current 
position of the hypothesised dark 
matter subhalo core at $(\ell, b)=(38.5^{\circ},-13.4{^\circ})$.
This final step, in practice, 
assumes that the subhalo has retained its integrity during the
orbit, which the simulations suggest is likely.  The data set is analysed using the binned likelihood method, 
implemented as the \texttt{gtlike} tool in the 
standard {\it Fermi} Science Tools\footnote{\tiny http://fermi.gsfc.nasa.gov/ssc/data/analysis/scitools/binned\_likelihood\_tutorial.html}.
At the position of the Smith cloud, we create a count map 
within our region of interest (ROI) with \texttt{gtbin}.   
We then generate a binned exposure map with  \texttt{gtexpcube2}, and  
a model source/diffuse 
count map with \texttt{gtsrcMaps}. In order to place an upper limit for
the integrated photon flux, 
we take advantage of the implementation of 
\texttt{LATAnalysisScripts}\footnote{\tiny http://fermi.gsfc.nasa.gov/ssc/data/analysis/scitools/LATAnalysisScripts.html}. 
 Upper limits
are computed with \texttt{calcUpper} assuming a power law spectrum
with photon index $\Gamma = 2$.
The 95\% confidence upper limit in the 1--300~GeV energy range
is $1.4\times 10^{-10}$~ph~cm$^{-2}$~s$^{-1}$.

\section{Bounds on dark matter cross sections}
\label{classp}
Our next task is to translate the derived {\it Fermi} upper limits into
actual constraints on annihilation cross sections. 
In order to so, we use the predicted gamma-ray flux
from dark matter annihilation \citep{Baltz2008} that can  be written as  

\begin{equation}
     \Phi_{\gamma}(E_{\gamma},\Delta\Omega)
        = \Phi^{\rm pp}(E_{\gamma}) \times J(\Delta\Omega)\,,
\label{annihilation}
 \end{equation}

\noindent
where $\Phi^{\rm pp}(E_{\gamma})$ is the
``particle physics factor'' that measures the number of gamma-ray photons
per dark matter annihilation. The quantity 
$J$ is the ``astrophysical factor'' which depends on the
integral of the dark matter distribution $\rho_{\rm DM}$ 
over the line of sight $l$. According to this model,

 \begin{equation}
      J = \int_{\Delta\Omega}\int \rho_{\rm DM}^2 (l,\Omega)\,\d{}l\d{}\Omega.
      \label{eq:J}
 \end{equation}

Here, the solid angle $\Delta\Omega$ as a function of 
the integration angle $\alpha_{\rm int}$
is defined as

\begin{equation} 
  \Delta\Omega = 2\pi\cdot(1-\cos(\alpha_{\rm int})) \,.
\end{equation}

As emphasized previously, 
since we do not know actually know where in the orbit 
of Smith the dark-matter subhalo is
located, we are making the simplified assumption -- supported by simulations --  that it
is still mostly intact at the head of the  cometary structure after 
its passage of the Galactic disc. We can then estimate $J$ assuming
the values for the dark matter distribution from Section 2.  
All calculations of $J$ presented therein have been obtained with 
the \texttt{CLUMPY} package \citep{clumpy}. 
For the dark-matter subhalo, we consider a canonical 
NFW profile 
\citep{nfw},

\begin{equation}
  \rho_{\rm NFW}(r)=\rho_s\biggl (\frac{r}{r_s}\biggr )^{-1}\biggl [1+\biggl (\frac{\
r}{r_s}\biggr )\biggr ]^{-2}.
  \label{nfw}
\end{equation}

where $\rho_s$ is the scale density, and  $r_s$ is the scale radius.
 For the actual computation, we adopt
$\rho_s=1.5\times10^{7}$~M$_\odot$~kpc$^{-3}$ and $r_s=1.04$~kpc from
\citet{Nichols2009}, a value approximately equal to the dark matter halo in numerical simulations. Integrating with 
$\alpha_{\rm int}=0.1^\circ$ (used throughout the text) 
which is compatible with the {\it Fermi}-LAT PSF at energies $>$ 1 GeV, 
we find  $\log_{10} J_{\rm NFW}$ = 18.44~GeV$^2$~cm$^{-5}$.

To allow for a different spatial model, 
we can also calculate the dark matter distribution with 
an Einasto profile \citep{einasto,Springel2008} of the form, 

\begin{equation}
\rho_{\rm E}(r)=\rho_{-2}\exp\left\{-\frac{2}{\alpha}\;\left[\left(\frac{r}{r_{-2}}\right)^{\alpha}-1\right]\right\},
\label{einasto}
\end{equation}

where $r_{-2}$ marks the radius where the slope of the profile
equals the isothermal value $\gamma = 2$, and $\alpha$$\sim$$0.18$ is the Einasto parameter.
We assume the scalings provided by \citet{navarro} so that
$r_{-2} = r_{d}$, and $\rho_{s} = 4 \rho_{-2}$ respectively. This
increases the value of the astrophysical value slightly
to $\log_{10} J_{\rm E}$ = 18.52 GeV$^2$~cm$^{-5}$.

\begin{figure}                                                                

\hfil
\includegraphics[width=3.1in,angle=0.]{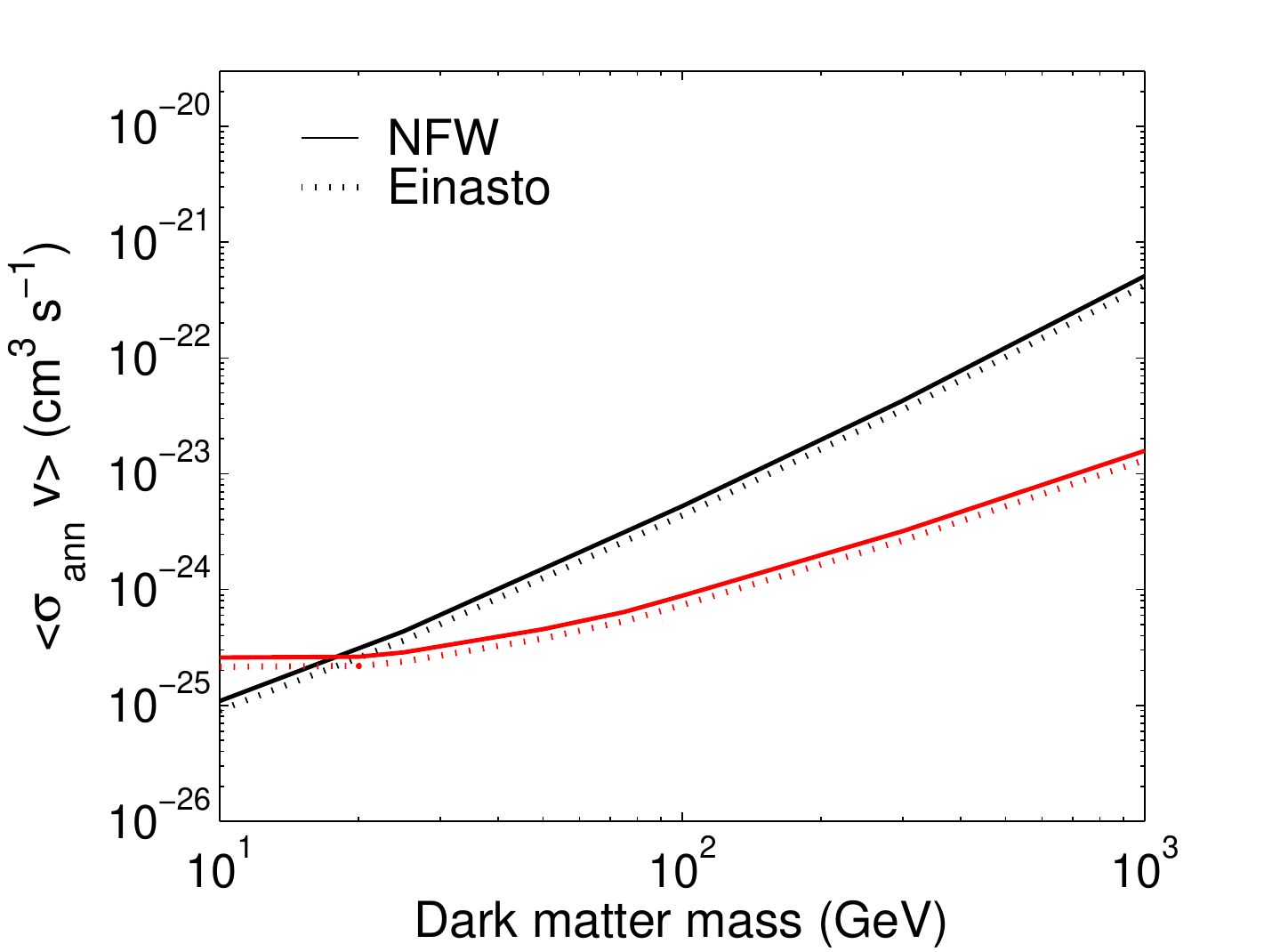}
\hfil
\caption{Derived upper limits for 
$\langle\sigma v\rangle_{\chi}$ versus dark matter mass
for both the NFW and Einasto profiles. 
$\tau^{+}{\tau}^{-}$ (black) and $b\bar{b}$ (red) channels are shown.
}
\label{figure2}
\end{figure}

Referring back to \citet{Baltz2008}, we can write the particle physics term 
$\Phi^{\rm pp}(E_{\gamma})$ as

\begin{equation}
     \Phi^{\rm pp}(E_{\gamma})\equiv  \frac{\mathrm{d}\Phi_{\gamma}}{\mathrm{d}E_{\gamma}}
        = \frac{1}{4\pi}\frac{\langle\sigma v\rangle}{2m_{\chi}^{2}}
          \times \sum_{f} \frac{\d{}N_{\gamma}}{\d{}E_{\gamma}}B_f,
 \end{equation}

where $\langle\sigma v\rangle_{\chi}$ is the
averaged annihilation cross section times the relative velocity,
$m_{\chi}$ is the dark matter particle mass, 
$\mathrm{d}N_{\gamma}/\mathrm{d} E_{\gamma}$ is the 
photon yield per annihilation final state $f$, and $B_{f}$ is the branching
ratio. For our calculations of $\mathrm{d}N_{\gamma}/\mathrm{d} E_{\gamma}$, 
we use the analytical fits to \texttt{PYTHIA} simulation spectra provided in 
 \citet{fornengo}.  We only consider dark matter annihilating into  
$\tau$-lepton pair ($\tau^{+}{\tau}^{-}$) and bottom quark 
($b\bar{b}$) final states. 
Fig. \ref{figure2} shows the cross section upper limits as a function of dark matter
mass for both channels.

\section{Discussion and conclusions}
\label{interp}

We have undertaken analytic calculations and numeric simulations to investigate whether the Smith Cloud is encapsulated within a dark matter halo.
In the absence of dark matter, only the most dense and correspondingly massive clouds survive the passage through the disc of the Galaxy.
Such massive objects show column densities much higher than the Smith Cloud is observed to have, and morphologically consist of a dense core with connected tendrils of gas.
With a dark matter halo, a HVC is able to survive the passage through substantially intact.
At low densities, $n\sim0.2$~cm$^{-3}$, the resulting cloud contains a column density similar to that of the Smith Cloud and a similar morphological structure to the Smith Cloud of fragmented clumps unattached to the main structure but following the orbit.
Such simulations suggest that the idea of the Smith Cloud being encapsulated by a dark matter halo remains plausible, in particular for this orbit, morphology and $N_{\rm HI}$ distribution.

For more massive clouds, the inner densities are approaching that of star forming regions in metal poor dwarf galaxies \citep{Ekta2008}. 
The spatial segregation of gas free dSphs and dIrrs, where the former is mostly found close to the main galaxy, in the Milky Way and Andromeda is evidence for efficient gas stripping closer to the halo centres \citep{Grcevich2009}. Any star formation would lead to supernova feedback greatly assisting the stripping of gas from the HVC \citep{Gatto2013}, regardless of the presence of dark matter. In light of \cite{Gatto2013}, the absence of stellar feedback in the Smith Cloud makes the fact that we see such a gas rich structure, which possibly is an "unformed" dwarf galaxy,  less of a timing problem, as the compact gas cloud may survive many disc crossings without being completely destroyed via ram pressure/tidal stripping. To fully understand the plausibility of relating an object like the Smith Cloud to non-star forming substructure, a fully cosmological context is necessary.

If encapsulated by dark matter, the Smith Cloud therefore inhabits a narrow region between being too light to survive ram pressure removing it from the host dark matter subhalo and being too massive with resulting star formation greatly assisting this stripping. 

We report photon flux upper limits for the gamma-ray emission at the current
position of 
the Smith Cloud using 
4.85 years of accumulated 
{\it Fermi}-LAT data to investigate the properties of such a dark matter halo. 
We exclude WIMPs annihilating 
into $\tau^{+}{\tau}^{-}$ and $b\bar{b}$ final states down to
$\langle\sigma v\rangle_{\chi} \sim 3\times 10^{-25}$ cm$^{3}$ s$^{-1}$
for masses around 10 GeV.
 The obvious caveat is that
dark matter might have been shed during the history of the cloud, however, the simulations have the dark matter retain the cloud core up to its present day location with only minor elongation. 
To investigate this case we compare the counts from the projected orbit and anti-orbit.
We find no evidence for excess gamma-ray emission 
along the predicted trajectory of the cloud system. In addition,
there is no morphological 
gamma-ray structures overlapping the cometary structure reported in \HI{}.
Despite this failed effort, gamma-ray observations still offers one of the
few available opportunities to diagnose the 
dark matter content of gaseous clouds. 
Upcoming experiments such as the Cherenkov Telescope Array (CTA) will be able to achieve improved angular resolution and sensitivity 
around the Smith Cloud 
for energies above 100~GeV 
\citep{cta,doro}.

The velocity averaged annihilation cross section upper bounds obtained 
around the Smith cloud are compatible with limits from other searches 
reported in the dwarf galaxies \citep{dwarfs} and galaxy
clusters \citep{clusters}.
Our results rest upon the assumption that there is indeed
a dark-matter subhalo seeding
the Smith Cloud.
To this extent we undertook simulations demonstrating that such an assumption is plausible given the disc passage the Smith Cloud is likely to have undertaken.
As well as demonstrating that the Smith Cloud's structure can be reproduced through a dark matter embedded HVC, such a result suggests that clouds with dark matter that have passed a disc may take on the comet like morphology observed in expected in DM free HVCs which have not yet crossed the disc \citep[\eg][]{Putman2011,Plockinger2012}.
The latter still needs 
to be verified with future observations, however, it provides potentially the best candidate for a dark matter confined HVC.

\section*{Acknowledgments}
N.M. acknowledges support from the Spanish taxpayers 
through a Ram\'on y Cajal fellowship and the 
Consolider-Ingenio 2010 Programme under grant MultiDark CSD2009-00064. 
All numerical simulations were conducted on the RCC Midway cluster at the University of Chicago.
OA is grateful to Doug Rudd for making the use of the Midway cluster a smooth experience.
We thank Tarek Hassan for his help with general technicalities.

\label{lastpage}
\end{document}